\documentclass[preprint]{revtex4}

\usepackage{graphicx}
\usepackage{dcolumn}
\usepackage{amsmath}
\usepackage{amsfonts}

\usepackage[T1]{fontenc}
\usepackage[latin1]{inputenc}
\usepackage{booktabs}
\usepackage[font=small,labelfont=bf,tableposition=top]{caption}
\usepackage{rotating}
\usepackage{array,multirow}
\usepackage{float}

\begin{document}

\title{Lie symmetry analysis of the Einstein-Maxwell equations for quark stars}

\author{M. A. Z. Khan}\email[]{muhammadalzafark@gmail.com}
\author{R. B. Narain}\email[]{narain@ukzn.ac.za}

\affiliation{Astrophysics and Cosmology Research Unit, School of Mathematics, Statistics and Computer Science, 
University of KwaZulu--Natal, Private Bag X54001, Durban 4000, South Africa}

\begin{abstract}
We derive the Lie point symmetries for the MIT Bag Model for quark stars in relativistic astrophysics. Four cases of reduction arise; three cases of specific values of the meausre of the anisotropy variation, and one general case, which we postulate as a specific relationship between the two gravitational potentials. We demonstrate the applicability of the model by generating two closed form solutions that satisfy the master gravitational equation and we match the interior geometries of the gravitating hyperspheres with the external solution given by the Reissner-Nordstr$\ddot{\text{o}}$m metric at the stellar boundary. Lastly, we produce a general class of solutions that are attainable for smooth and continuous functions and generate two exact solutions using this model. \\ \\
\emph{Keywords}: Nonlinear equations; Lie symmetries; Quark stars, Einstein-Maxwell field equations 
\end{abstract}

\maketitle


\section{Introduction}
The Einstein-Maxwell equations are a system of nonlinear ordinary differential equations with two dependent variables, the gravitational potentials, and one independent variable, the radial coordinate. Complications in solving these equations arise due to the radial evolution of the electric field intensity. Further impediments befall the model in the context of pressure anisotropy, where the radial and tangential pressures are unequal, in general, and their absolute difference is allowed to vary radially. 

The study of the solutions to these equations are pre-eminent in relativistic astrophysics which arise in the study of compact objects, in particular quark stars, gravastars, neutron stars, dark energy stars and black holes. The solutions to these equations have been studied extensively for various situations; Ivanov \cite{Ivanov} investigated classification schemes for perfect fluid hyperspheres and thereafter matched the resultant interior geometries attained with the exterior Reissner-Nordstr$\ddot{\text{o}}$m solution \cite{Jeffery}, \cite{Weyl}, \cite{Nordstrom} and \cite{Reissner} at the stellar surface. 

Sharma and Maharaj \cite{SharmaMaharaj} established new results for compact stars by specifying a form for the mass function and it was shown that linear equations of state were manifest due to the nature of the solutions attained; thus being congruous in the description of quark matter and strange stars. 

Thirukkanesh and Maharaj \cite{ThirukkaneshMaharaj} found three new solutions to the field equations by specifying forms for one of the gravitational potentials and the electric field intensity. The physical viability of these solutions were demonstrated by generating plots for the energy density, radial and tangential pressures, electric field intensity and measure of the anisotropy. 

Thirukkanesh and Ragel \cite{ThirukkaneshRagel} studied compact stars and quark matter in the context of a linear equation of state; in particular they chose the constants from the MIT Bag Model \cite{Chodus} and \cite{DeGrand}. They matched the interior geometry attained with the Schwarzchild exterior solution \cite{SchwarzchildA} and \cite{SchwarzchildB} at the stellar boundary and then proceeded to show that all physical parameters were satisfied.

Mafa Takisa and Maharaj \cite{MafaTakisaMaharaj} produced new solutions to the field equations by choosing a barotropic equation of state. A salient feature exhibited in this model was how the electric field intensity affected the mass of charged stellar objects. Lastly, the applicability and physical feasibility of this model was displayed by re-attaining the observed mass of a binary pulsar.

Other noteable work in this area has been carried out by Bijalwan \cite{Bijalwan}, Negreiros \textit{et. al.} \cite{NegreirosWeberMalheiroUsov} and Murad and Fatema \cite{MuradFatema} who studied charged models for the case of pressure isotropy. Esculpi and Aloma \cite{EsculpiAloma} studied the general case of charged models with pressure anisotropy. 

The foremost objective of this study is to establish a relation between the gravitational potentials with a particular case of the linear equation of state, \textit{viz.} the MIT Bag Model \cite{Chodus} and \cite{DeGrand}, and find new classes of solutions based on this relation. This aim is achieved by employing the use of Lie symmetry techniques for differential equations, see \cite{LieA} and \cite{LieB}. The application of Lie symmetry methods to nonlinear ordinary and partial differential equations that arise in relativistic astrophysics is not novel; several researchers have contributed in this respect.  

Maharaj \textit{et. al.} \cite{MaharajLeachMaartens} studied the integrability properties of the field equations for spherically symmetric shear-free geometries. First integrals were obtained and a recovery of known solutions was performed showing that all other cases investigated were special cases of the solutions they had attained.

Kweyama \textit{et. al.} \cite{KwyamaMaharajGovinderA}, \cite{KwyamaMaharajGovinderB} and \cite{KwyamaMaharajGovinderC} studied perfect and charged fluid hyperspheres by finding first integrals, Lie point symmetries and Noether symmetries of the gravitational potential functions in the field equations.

Abebe \textit{et. al.} \cite{AbebeMaharajGovinderA}, \cite{AbebeMaharajGovinderB}, \cite{AbebeMaharajGovinderC} and \cite{AbebeMaharajGovinderD} studied the radial and time evolutions of the nonlinear field equations in the context of conformally flat stars with radiation. The Lie point symmetries attained were used to derive an optimal system of symmetries. Using the Lie symmetry approach, the generation of geodesic models were outlined. Lastly, the more general situation of a relativistic star with expansion, shear and acceleration was investigated and new exact solutions were found, thus demonstrating the far reaching capabilities of the Lie symmetry approach.

Other noteworthy investigations in this area has been carried out by Mohanlal \textit{et. al.} \cite{MohanlalMaharajTiwariNarain} and Maharaj \textit{et. al.} \cite{MaharajTiwariMohanlalNarain} who studied spherical bodies with radiation. Msomi \textit{et. al.} \cite{MsomiGovinderMaharajA}, \cite{MsomiGovinderMaharajB} and \cite{MsomiGovinderMaharajC} studied shear free spherically symmetric spacetimes in higher dimensions, models with heat flow, embedding class one fluids and spacetimes with conformally related geometries.

In \S II, we glean the Einstein-Maxwell field equations, apply a transformation, for the purposes of simplification, upon them and thereafter derive the underlying master gravitational equation. In \S III, we explore a solution strategy at solving for the potentials in the Einstein-Maxwell equations by rearranging the relation for the electric field intensity and integrating. In this manner, several new classes of solutions can be obtained provided integrable forms of electric field intensity can be chosen and the other gravitational potential. In \S IV, we determine Lie point symmetries of the underdetermined master gravitational equations rigorously by solving the resulting set of linear partial differential equations. In \S V we identify four categories of solutions that can arise for specific forms of the measure of the pressure anisotropy and tabulate the reduction procedures. With the most general form of the measure of the pressure anisotropy, we state a general result between the two gravitational potentials. In \S VI, we demonstrate the capabilities of the results obtained in \S V by identifying three classes of new solutions that can be easily deduced; the third case we state a general result for general functions. In section \S VII, we deliberate the results produced in this paper and consider possible future pursuits.

\section{The model}\label{sect2}
In this section we derive the Einstein-Maxwell equations in full. We apply a transformation upon them for the purposes of unembellishment. Lastly we rearrange the relation for the measure of the anisotropy in order to produce the master gravitational equation.

The interior geometry is described by the static spherically symmetric metric
\begin{eqnarray}\label{1}
\text{d}s^{2}_{-}=-\text{e}^{2\nu}\text{d}t^{2}+\text{e}^{2\lambda}\text{d}r^{2}+r^{2}(\text{d}\theta^{2}+\sin^{2}\theta\;\text{d}\phi^{2}),
\end{eqnarray}
where $\nu=\nu(r)$ and $\lambda=\lambda(r)$ are the gravitational potential functions. The exterior geometry is described by the Reissner-Nordstr$\ddot{\text{o}}$m metric \cite{Jeffery}, \cite{Weyl}, \cite{Nordstrom} and  \cite{Reissner}
\begin{eqnarray}\label{2}
\text{d}s^{2}_{+}=-\left[1-\frac{2M}{r}+\left(\frac{Q}{r}\right)^{2}\right]\text{d}t^{2}+\left[1-\frac{2M}{r}+\left(\frac{Q}{r}\right)^{2}\right]^{-1}\text{d}r^{2}+r^{2}(\text{d}\theta^{2}+\sin^{2}\theta\;\text{d}\phi^{2}),
\end{eqnarray}
where $M$ and $Q$ are the total mass and charge measured by an observer at an infinite distance away respectively. In order to describe anisotropic charged matter, we consider the form of the energy-momentum tensor
\begin{eqnarray}\label{(3)}
T_{ij}=\text{diag}\left[-\left(\rho+\frac{E^{2}}{2}\right),p_{\parallel}-\frac{E^{2}}{2},p_{\perp}+\frac{E^{2}}{2},p_{\perp}+\frac{E^{2}}{2}\right],
\end{eqnarray}
where $\rho$ denotes the energy density, $p_{\parallel}$ denotes the radial pressure, $p_{\perp}$ denotes the tangential pressure and $E$ denotes the local electric field intensity. These physical quantities are measured relative to a unit and timelike vector $\mathbf{v}$ that has comoving components $v^{j}$, and $j$ is the running index in the set $\{0,1,2,3\}$. The modified Einstein field equations, the Einstein-Maxwell equations, due to the incorporation of the local electric field, has the form
\begin{eqnarray}\label{4}
\begin{split}
&\frac{1-\text{e}^{-2\lambda}}{r^{2}}+\frac{2\lambda'\text{e}^{-2\lambda}}{r}=\;\rho+\frac{E^{2}}{2},\\
&-\frac{1-\text{e}^{2\lambda}}{r^{2}}+\frac{2\nu'\text{e}^{-2\lambda}}{r}=\;p_{\parallel}-\frac{E^{2}}{2},\\
&\text{e}^{-2\lambda}\left[\nu''+(\nu')^{2}-\nu'\lambda'+\frac{1}{r}(\nu'-\lambda')\right]=\;p_{\perp}+\frac{E^{2}}{2},\\
&\sigma=\;\frac{\text{e}^{-\lambda}(r^{2}E'+2rE)}{r^{2}},
\end{split}
\end{eqnarray}
where the primes $(')$ denote differentiation with respect to the radial coordinate $r$ and $\sigma$ denotes the proper charge density. In the Einstein-Maxwell equations, we have adopted the convention of setting the speed of light $c$ and the guage (coupling) constant $\frac{8\pi G}{c^{4}}$ to unity. Within a static spherically symmetric charged gravitating hypersphere, the mass is given by
\begin{eqnarray}\label{5}
m(r)=\frac{1}{2}\int_{0}^{r}\;\omega^{2}(\tilde{\rho}+E^{2})\;\text{d}\omega,
\end{eqnarray}
where $\tilde{\rho}=\rho|_{E=0}$, that is to say, $\tilde{\rho}$ is the energy density when the local electric field is zero. For quark matter, the relationship between the energy density and radial pressure is of the form
\begin{eqnarray}
p_{\parallel}=\frac{1}{3}(\rho-4B),\label{(6)}
\end{eqnarray}
where $B$ is the bag constant arising from the MIT Bag Model used to describe relativistic quarks - hadrons - see \cite{Chodus} and \cite{DeGrand}. We define the transformations
\begin{eqnarray}\label{7}
\begin{split}
x:=&\;Cr^{2},\\
A^{2}y^{2}:=&\;\text{e}^{2\nu},\\
z:=&\;\text{e}^{-2\lambda},
\end{split}
\end{eqnarray}
that was introduced by Durgapal and Bannerji \cite{BannerjiDurgapal}, where $x$ is the new independent variable, $y=y(x)$ and $z=z(x)$ are the new gravitational potential functions and $A, C\in\mathbb{R}\backslash\{0\}$ are arbitrary constants. The use of this set of transformations are vital for the simplification of the field equations \eqref{4} and has been used triumphantly in the study of Einsteinian and modified Einstein theories of gravity; for example, Maharaj \textit{et. al.} \cite{MaharajChilambweHansraj} and Hansraj \textit{et. al.} \cite{HansrajChilambweMaharaj} employed the use of the transformation \eqref{7} in the study of Einstein-Gauss-Bonnet gravity and more recently, Dadhich \textit{et. al.} \cite{DadhichHanrajChilambwe} utilized \eqref{7} in the study of pure Lovelock gravity \cite{Lovelock}. Upon application of the transformation \eqref{7} on \eqref{1} and \eqref{5}, the interior metric takes the form
\begin{eqnarray}
\text{d}s^{2}_{-}=-A^{2}y^{2}\text{d}t^{2}+\frac{\text{d}x^{2}}{4xCz}+\frac{x}{C}(\text{d}\theta^{2}+\sin^{2}\theta\;\text{d}\phi^{2}),
\end{eqnarray}
and the mass function becomes
\begin{eqnarray}
m(x)=\frac{1}{C^{3/2}}\int_{0}^{x}\;\sqrt{\omega}(\tilde{\rho}+E^{2})\;\text{d}\omega,
\end{eqnarray}
where $\tilde{\rho}$ takes the new form
\begin{eqnarray}
\tilde{\rho}=C\left(\frac{1-z}{x}-2\dot{z}\right),
\end{eqnarray}
and the overdot notation $(\dot{})$ denotes differentiation with respect to $x$. Additionally with the transformations (7), the Einstein-Maxwell equations (4) take the form
\begin{eqnarray}\label{10}
\begin{split}
\rho=&\;3p_{\parallel}+4B,\\
\frac{p_{\parallel}}{C}=&\;\frac{\dot{y}z}{y}-\frac{\dot{z}}{2}-\frac{B}{C},\\
p_{\perp}=&\;p_{\parallel}+\Delta,\\
\Delta=&\;\frac{4xC\ddot{y}z}{y}+\frac{C\dot{y}(2x\dot{z}+6z)}{y}+C\left[2\left(\dot{z}+\frac{B}{C}\right)+\frac{z-1}{x}\right],\\
\frac{E^{2}}{2C}=&\;\frac{1-z}{x}-\frac{3\dot{y}z}{y}-\frac{\dot{z}}{2}-\frac{B}{C},\\
\sigma=&\;2\sqrt{\frac{Cz}{x}}(x\dot{E}+E),
\end{split}
\end{eqnarray}
where $\Delta$ is the measure of the pressure anisotropy and is defined simply as $\Delta=p_{\perp}-p_{\parallel}$. We observe that the system of equations \eqref{10} has six equations with eight unknown variables $\{\rho,p_{\parallel},p_{\perp},\Delta,E,\sigma,y,z\}$. Therefore, in order to solve the system of equations exactly, two quantities need to be specified. Sunzu \textit{et. al.} \cite{SunzuMaharajRay2} and \cite{SunzuMaharajRay1} and Maharaj \textit{et. al.} \cite{SunzuMaharajRay3} solved the system \eqref{10} by specifying forms for the measure of the anisotropy $\Delta$ and the potential $y$ and thus demonstrating the physical reliability of these choices. We resort to solving the equations exactly by using the Lie symmetry approach. In equation \eqref{10}, we rearrange the relationship for the measure of the pressure anisotropy in order to obtain the master gravitational equation
\begin{eqnarray}\label{11}
4Cx^{2}z\ddot{y}+(6Cxz+2Cx^{2}\dot{Z})\dot{y}+(2Bx+Cz+2Cx\dot{z}-x\Delta-C)y=0.
\end{eqnarray}
 Equation \eqref{11} shall be the objective of our analysis hereof and we shall apply the method of Lie symmetries in order to obtain point symmetries that allow for the mapping of every solution to a group orbit of other solutions, see the work of Ibragimov \cite{ibragimov}.
 

\section{ANALYSIS}
In this section we create a relationship between the gravitational potentials $y$ and $z$ by rearranging the form of the equations for $\frac{E^{2}}{2C}$. In this way, we give an elementary expression, in terms of integrating factors, for $y$.

Rearranging the equations for $\frac{E^{2}}{2C}$ in \eqref{10}, we get
\begin{eqnarray}\label{12}
6Cxz\dot{y}+[Cx\dot{z}+2Cz+xE^{2}(x)+2Bx-2C]y=0.
\end{eqnarray}
Integrating \eqref{12}, we get
\begin{eqnarray}\label{15}
y(x)=&\;y_{0}\exp\left[-\frac{1}{6C}\int\;\left(Cx\dot{z}+2Cz+xE^{2}(x)+2Bx-2C\right)\;\text{d}x\right],
\end{eqnarray}
where $y_{0}$ is the constant of integration. Equation \eqref{15} allows one to specify any form for the electric field intensity $E$ and, for a suitable form of $z$, obtain the gravitational potential $y$. In \S\ref{Lie Symmetries}, we demonstrate a similar procedure by employing the method of Lie symmetries to obtain a relationship between $y$ and $z$.


\section{Lie Symmetries}\label{Lie Symmetries}
We consider Lie point symmetries of the form
\begin{eqnarray}\label{20}
\Gamma=\xi(x,y,z)\partial{x}+\eta(x,y,z)\partial{y}+\varphi(x,y,z)\partial_{z}.
\end{eqnarray}
The prolongation operator of \eqref{20}, up to second order for $y$, is given by
\begin{eqnarray}\label{13}
\Gamma^{[2]}=\Gamma+\zeta^{[1],y}\partial_{\dot{y}}+\zeta^{[1],z}\partial_{\dot{z}}+\zeta^{[2],y}\partial_{\ddot{y}},
\end{eqnarray}
where
\begin{eqnarray}\label{14}
\begin{split}	
\zeta^{[1],y}=&\;\eta_{x}+(\eta_{y}-\xi_{x})\dot{y}+\eta_{z}\dot{z}-\xi_{y}(\dot{y})^{2}-\xi_{z}\dot{y}\dot{z},\\
\zeta^{[1],z}=&\;\varphi_{x}+(\varphi_{z}-\xi_{x})\dot{z}+\varphi_{y}\dot{y}-\xi_{z}(\dot{z})^{2}-\xi_{y}\dot{y}\dot{z},\\
\zeta^{[2],y}=&\;\eta_{xx}+(2\eta_{xy}-\xi_{xx})\dot{y}+(\eta_{yy}-2\xi_{xy})(\dot{y})^{2}-\xi_{yy}(\dot{y})^{3}+(\eta_{y}-2\xi_{x}-3\xi_{y}\dot{y})\ddot{y}\\
&-2[\xi_{z}\ddot{y}+\xi_{yz}(\dot{y})^{2}+(\xi_{xz}-\eta_{yz})\dot{y}-\eta_{xz}]\dot{z}+(\eta_{zz}-\xi_{zz}\dot{y})(\dot{z})^{2}+(\eta_{z}-\xi_{z}\dot{y})\ddot{z}.
\end{split}
\end{eqnarray}
Direct application of Lie's invariant surface condition requires the praxis of \eqref{13} and \eqref{14} on \eqref{11} together with the restriction
\begin{eqnarray}
\ddot{y}=\frac{(C-2Bx-Cz+x\Delta-2Cx\dot{z})y-2Cx(x\dot{z}+3z)\dot{y}}{4Cx^{2}z}.
\end{eqnarray}
For a full elucidation of the theory of Lie symmetries of differential equations, the reader is encouraged to see the works of Ibragimov \cite{ibragimov}, Olver \cite{olver} and Stephani \cite{stephani}. Separating by powers of the derivatives $\dot{y}$ and $\dot{z}$, we get the set of determining equations
\begin{eqnarray}\label{16}
\begin{split}
(\dot{y})^{3}:&\;-4Cx^{3}z^{2}\xi_{yy}=0,\\
(\dot{y})^{2}:&\;2Cx^{3}z\varphi_{y}+4Cx^{3}z^{2}\eta_{yy}-8Cx^{3}z^{3}\xi_{xy}+12Cx^{2}z^{2}\xi_{y}=0,\\
(\dot{y})^{2}\dot{z}:&\;2Cx^{3}z\xi_{y}=0,\\
\dot{y}:&\;2Cx^{3}z\varphi_{x}-3Cxyz\xi_{y}+6Bx^{2}yz\xi_{y}+2Cx^{2}yz\varphi_{y}-6Cxz^{2}\xi+8Cx^{3}z^{2}\eta_{xy}\\
&+6Cx^{2}z^{2}\xi_{x}-4Cx^{3}z^{2}\xi_{xx}+3Cxyz^{2}\xi_{y}-3x^{2}yz\Delta\xi_{y}=0,\\
\dot{y}\dot{z}:&\;2Cx^{3}z\varphi_{z}+4Cx^{2}yz\xi_{y}-2Cx^{3}\varphi-6Cx^{2}z^{2}\xi_{z}=0,\\
\dot{y}(\dot{z})^{2}:&\;-4Cx^{3}z\xi_{z}=0,\\
(\dot{y})^{0}:&\;Cxy\varphi-2Bx^{2}y\varphi-Cxz\eta+2Cx^{2}z\eta+2Cyz\xi+Cxyz\eta_{y}-2Bx^{2}yz\eta_{y}\\
&-2Cxyz\xi_{x}+4Bx^{2}yz\xi_{x}+2Cx^{2}yz\varphi_{x}+Cxz^{2}\eta+6Cx^{2}z^{2}\eta_{x}+4Cx^{3}z^{2}\eta_{xx}\\
&-2Cyz^{2}\xi-Cxyz^{2}\eta_{y}+2Cxyz^{2}\xi_{x}+x^{2}y\Delta\varphi-x^{2}z\Delta\eta+xyz\Delta\xi+x^{2}yz\Delta\eta_{y}\\
&-2x^{2}yz\Delta\xi_{x}=0,\\
\dot{z}:&\;2Cx^{2}z\eta-2Cx^{2}y\varphi+2Cx^{3}z\eta_{x}-2Cxyz\xi-2Cx^{2}yz\eta_{y}+2Cx^{2}yz\xi_{x}+2Cx^{2}yz\varphi_{z}\\
&+6Cx^{2}z^{2}\eta_{z}=0,\\
(\dot{z})^{2}:&\;2Cx^{3}z\eta_{z}-2Cx^{2}yz\xi_{z}=0.
\end{split}
\end{eqnarray}
The system \eqref{16} is a set of linear partial differential equations. Solving  \eqref{16}, we get four cases of Lie point symmetries that arise; three sets of symmetries for three specific choices of $\Delta$ and one general case for $\Delta$, which is $\Delta=\Delta(x)$.  

\subsection{Case 1. $\Delta=0$}
For the specific choice of $\Delta=0$, \eqref{11} becomes the isotropic relation
\begin{equation}\label{21}
4Cx^{2}z\ddot{y}+(6Cxz+2Cx^{2}\dot{z})\dot{y}+(2Cx\dot{z}+2Bx+Cz-C)y=0.
\end{equation}
Equation \eqref{21} admits the one-dimensional subalgebra 
\begin{eqnarray}
\Gamma_{1}=y\partial_{y}.
\end{eqnarray}

\subsection{Case 2. $\Delta=2B$}
For the specific choice of $\Delta=2B$, the master equation \eqref{11} reduces to
\begin{eqnarray}\label{19}
4Cx^{2}z\ddot{y}+(6Cxz+2Cx^{2}\dot{z})\dot{y}+(2Cx\dot{z}+Cz-C)y=0.
\end{eqnarray}
Equation \eqref{19} admits the one-dimensional subalgebra
\begin{eqnarray}
\Gamma_{2}=x\partial_{x}.
\end{eqnarray} 

\subsection{Case 3. $\Delta=\frac{2Bx-C}{x}$} 
For the specific choice of $\Delta=\frac{2Bx-C}{x}$, the master equation \eqref{11} reduces to
\begin{eqnarray}\label{17}
4Cx^{2}z\ddot{y}+(6Cxz+2Cx^{2}\dot{z})\dot{y}+(2Cx\dot{z}+Cz)y=0.
\end{eqnarray}
Equation \eqref{17} admits a two-dimensional subalgebra of symmetries spanned by the vector fields
\begin{eqnarray}
\begin{split}
\Gamma_{3}=&\;y\partial_{y},\\
\Gamma_{4}=&\;z\partial_{z},
\end{split}
\end{eqnarray} 
with the Lie bracket relation $[\Gamma_{3},\Gamma_{4}]_{LB}=0$. By the action of the algebras, we find that the group formed is Abelian and thus a reduction via $\Gamma_{3}$, $\Gamma_{4}$ or a linear combination of symmetries $a\Gamma_{3}+b\Gamma_{4}$, where $a, b\in\mathbb{R}$ are arbitrary constants, is possible.

\subsection{Case 4. $\Delta=\Delta(x)$}
For the general case where $\Delta=\Delta(x)$ where $\Delta$ varies arbitrarily, equation \eqref{19} remains unchanged and admits the one-dimensional subalgebra 
\begin{eqnarray}\label{23}
\Gamma_{5}=y\partial_{y}.
\end{eqnarray}
 
\section{Reductions}

In this section, we apply the Lie point symmetries generated for the various choices of $\Delta$ and produced closed form solutions for the potentials $y$ and $z$. We demonstrate the calculation of one class of solution, the case when $\Delta=\Delta(x)$, and thereafter state the results for the other cases in tabular form.

Using $\Gamma_{5}$, the Lie point symmetry \eqref{23} has the associated Lagrange system
\begin{eqnarray}\label{24}
\frac{\text{d}x}{0}=\frac{\text{d}y}{y}=\frac{\text{d}z}{0}=\frac{\text{d}\dot{y}}{\dot{y}}.
\end{eqnarray}  
From \eqref{24} we infer the relations
\begin{eqnarray}\label{25}
\begin{split}
u=&\;x,\\
P(u)=&\;y,\\
Q(u)=&\;z,\\
R(u)=&\;\frac{\dot{y}}{y},\\
\delta(u)=&\;\Delta.
\end{split}
\end{eqnarray}
Applying \eqref{25} to the master equation \eqref{11}, we get
\begin{eqnarray}\label{26}
R_{u}+\left[\frac{2uQP_{u}+P(uQ_{u}+3Q)}{2uPQ}\right]R+\left[\frac{(2CQ_{u}+2B-\delta)u+C(Q-1)}{4Cu^{2}PQ}\right]P=0.
\end{eqnarray}
Integrating equation \eqref{26}, we get
\begin{eqnarray}\label{27}
R(u)=\frac{1}{\sqrt{u^{3}Q}}\left[c_{1}-\frac{1}{4C}\int\;\frac{P[(2CQ_{u}-\delta+2B)u+C(Q-1)]}{\sqrt{uQ}}\;\text{d}u\right],
\end{eqnarray}
where $c_{1}$ is the constant of integration. In terms of the original coordinates, equation \eqref{27} can be expressed as
\begin{eqnarray}\label{28}
\frac{\dot{y}}{y}=\frac{1}{y\sqrt{x^{3}z}}\left[c_{1}-\frac{1}{4C}\int\;\frac{[2Cx\dot{z}+Cz+2Bx-x\Delta(x)-C]y}{\sqrt{xz}}\right]
\end{eqnarray}
We encapsulate the results of all reduction procedures in TABLE I below.
 
\begin{sidewaystable}[!htp]
\centering
\caption{Reduction variables, Reduced equations and solutions for various forms of the anisotropy $\Delta$}
\begin{tabular}{|p{3cm}|p{4cm}|p{6cm}|p{11cm}|}
\hline
\hline
\centering{\textbf{Form of Anisotropy}} & \centering{\textbf{Reduction Variables}} & \centering{\textbf{Reduced Equation}} & \hspace{4.5cm}\textbf{Solution} \\
\hline
$\Delta = 0$ & $x=u$ & $R_{u}+\left[\frac{2uQP_{u}+P(uQ_{u}+3Q)}{2uPQ}\right]$ & $y(x)=\int\;\frac{1}{\sqrt{x^{3}z}}\left[c_{1}-\frac{1}{4C}\int\;\frac{(2Cx\dot{z}+Cz+2Bx-C)y}{\sqrt{xz}}\;\text{d}x\right]\;\text{d}x+c_{2}$ \\
& $y=P(u)$ & $+\left[\frac{2(CQ_{u}+B)u+C(Q-1)}{4Cu^{2}PQ}\right]P=0$ & \\
& $z=Q(u)$ & & \\
& $\frac{\dot{y}}{y}=R(u)$ & & \\
\hline
$\Delta = 2B$ & $x=u$ & $R_{u}+\left(\frac{uQ_{u}+5Q}{2uQ}\right)R$ & $y(x)=\int\;\frac{1}{\sqrt{x^{3}z}}\left[c_{1}-\frac{1}{4}\int\;\frac{(2x\dot{z}+z-1)y}{\sqrt{xz}}\;\text{d}x\right]\text{d}x+c_{2}$ \\
& $y=P(u)$ & $+\left(\frac{C-2CuQ_{u}-CQ}{4Cu^{3}Q}\right)P=0$ & \\
& $z=Q(u)$ & & \\
& $-\frac{\dot{y}}{x}=R(u)$ & & \\ 
\hline
$\Delta = \frac{2Bx-C}{x}$ & $\Gamma_{3}: x=u$ & $R_{u}+\left[\frac{2uQP_{u}+P(uQ_{u}+3Q)}{2uPQ}\right]R$  & $y(x)=\int\;\frac{1}{\sqrt{x^{3}z}}\left[c_{1}-\frac{1}{4}\int\;\frac{(2x\dot{z}+z)y}{\sqrt{xz}}\;\text{d}x\right]\;\text{d}x+c_{2}$  \\
& $\;\;\;\;\;\;\;y=P(u)$ & $+\frac{2CuQ_{u}+CQ}{4Cu^{2}Q}=0$ & \\
& $\;\;\;\;\;\;\;z=Q(u)$ & & \\
& $\;\;\;\;\;\;\;\frac{\dot{y}}{y}=R(u)$ & & \\\cline{2-4}
& $\Gamma_{4}: x=u$ & $R_{u}+\left(\frac{uQ_{u}+3Q}{2uQ}\right)R$ & $y(x)=\int\;\frac{1}{\sqrt{x^{3}z}}\left[c_{1}-\frac{1}{4}\int\;\frac{(2x\dot{z}+z)y}{\sqrt{xz}}\;\text{d}x\right]\;\text{d}x+c_{2}$ \\
& $\;\;\;\;\;\;\;y=P(u)$ & $+\left(\frac{2uQ_{u}+Q}{4u^{2}Q}\right)P=0$ & \\
& $\;\;\;\;\;\;\;z=Q(u)$ & & \\
& $\;\;\;\;\;\;\;\dot{y}=R(u)$ & & \\
\hline
$\Delta = \Delta(x)$ & $x=u$ & $R_{u}+\left[\frac{2uQP_{u}+P(uQ_{u}+3Q)}{2uPQ}\right]R$ & $y(x)=\int\;\frac{1}{\sqrt{x^{3}z}}\left[c_{1}-\frac{1}{4C}\int\;\frac{[2Cx\dot{z}+Cz+2Bx-x\Delta(x)-C]y}{\sqrt{xz}}\;\text{d}x\right]\text{d}x+c_{2}$ \\
& $y=P(u)$ & $+\left[\frac{(2CQ_{u}+2B-\delta)u+C(Q-1)}{4Cu^{2}PQ}\right]P=0$ & \\
& $z=Q(u)$ & & \\
& $\frac{\dot{y}}{y}=R(u)$ & & \\ 
& $\Delta = \delta(u)$ & & \\ 
\hline
\end{tabular}
\end{sidewaystable}

\newpage
\textbf{Theorem 1:} In anisotropic Einstein-Maxwell gravity with the MIT Bag Model constants for the linear correspondence between energy density and radial pressure, the field equations admit a one-dimensional Lie algebra of symmetries $y\partial_{y}$. The gravitational potenials  $y$ and $z$ are interrelated by 
\begin{eqnarray}
y(x)=\int\;\frac{1}{\sqrt{x^{3}z}}\left[c_{1}-\frac{1}{4C}\int\;\frac{[2Cx\dot{z}+Cz+2Bx-x\Delta(x)-C]y}{\sqrt{xz}}\;\text{d}x\right]\;\text{d}x+c_{2},\nonumber
\end{eqnarray}
where $c_{1}$ and $c_{2}$ are constants of integration.



\section{PARTICULAR SOLUTIONS}
We demonstrate the generation of two closed form solutions of the Einstein-Maxwell equations by choosing specific forms for the measure of anisotropy from Theorem 1 in the previous section. We make the observation that solutions to the field equations are an inherent property of the possession of Lie point symmetries of the underlying master equation.

\subsection{Solution I}
For $\Delta=2B$, it is easy to verify that the potentials
\begin{eqnarray}\label{30}
\begin{split}
y(x)=&\;c_{2}-\frac{\sqrt{x}\sqrt{\sqrt{x}(\sqrt{x}+c_{1})}\sqrt{c_{1}\sqrt{x}+x}}{\sqrt{x^{5/2}(\sqrt{x}+c_{1})}},\\
z(x)=&\;\frac{c_{1}}{\sqrt{x}}+1.
\end{split}
\end{eqnarray} 
satisfy equation Theorem 1 above. Using the forms of the potential $y$ and $z$ in \eqref{30}, we generate the exact solution to the field equations \eqref{10}
\begin{eqnarray}
\begin{split}
\text{e}^{2\nu}=&\;A^{2}\left[c_{2}-\frac{\sqrt{x}\sqrt{\sqrt{x}(\sqrt{x}+c_{1})}\sqrt{c_{1}\sqrt{x}+x}}{\sqrt{x^{5/2}(\sqrt{x}+c_{1})}}\right],\\
\text{e}^{2\lambda}=&\;\frac{\sqrt{x}}{\sqrt{x}+c_{1}},\\
\rho=&\;\frac{3Cc_{1}c_{2}^{2}}{4x(c_{2}^{2}\sqrt{x}-\sqrt{x}-c_{1})}+\frac{3Cc_{1}c_{2}\sqrt{(\sqrt{x}+c_{1})x^{5/2}}}{4x^{5/2}(c_{2}^{2}\sqrt{x}-\sqrt{x}-c_{1})}+B,\\
p_{\parallel}=&\;\frac{Cc_{1}c_{2}^{2}}{4x(c_{2}^{2}\sqrt{x}-\sqrt{x}-c_{1})}+\frac{Cc_{1}c_{2}\sqrt{(\sqrt{x}+c_{1})x^{5/2}}}{4x^{5/2}(c_{2}^{2}\sqrt{x}-\sqrt{x}-c_{1})}-B,
\end{split}
\end{eqnarray}
\begin{eqnarray}
\begin{split}
p_{\perp}=&\;\frac{Cc_{1}c_{2}^{2}}{4x(c_{2}^{2}\sqrt{x}-\sqrt{x}-c_{1})}+\frac{Cc_{1}c_{2}\sqrt{(\sqrt{x}+c_{1})x^{5/2}}}{4x^{5/2}(c_{2}^{2}\sqrt{x}-\sqrt{x}-c_{1})}+B,\\
\Delta=&\;2B,\nonumber\\
E^{2}=&\;-\frac{3Cc_{1}c_{2}^{2}}{2x(c_{2}^{2}\sqrt{x}-\sqrt{x}-c_{1})}-\frac{3Cc_{1}c_{2}\sqrt{(\sqrt{x}+c_{1})x^{5/2}}}{2x^{5/2}(c_{2}^{2}\sqrt{x}-\sqrt{x}-c_{1})}-2B,\\
m(x)=&\;\frac{c_{1}\sqrt{x}+c_{1}^{2}}{\sqrt{C}(\sqrt{x}-c_{2}^{2}\sqrt{x}+c_{1})}-\frac{c_{1}c_{2}\sqrt{(\sqrt{x}+c_{1})x^{5/2}}}{\sqrt{C}x(c_{2}^{2}\sqrt{x}-\sqrt{x}-c_{1})}-\frac{4Bx^{3/2}+3Cc_{1}}{3C^{3/2}}.
\end{split}
\end{eqnarray}

Using \eqref{30} in \eqref{1}, the interior geometry is given by
\begin{eqnarray}\label{33}
\begin{split}
\text{d}s^{2}_{-}=&\;-A^{2}\left[c_{2}-\frac{\sqrt{x}\sqrt{\sqrt{x}(\sqrt{x}+c_{1})}\sqrt{c_{1}\sqrt{x}+x}}{\sqrt{x^{5/2}(\sqrt{x}+c_{1})}}\right]^{2}\text{d}t^{2}+\frac{\text{d}x^{2}}{4C\sqrt{x}(\sqrt{x}+c_{1})}\\
&+\frac{x}{C}(\text{d}\theta^{2}+\sin^{2}\theta\;\text{d}\phi^{2}).
\end{split}
\end{eqnarray}
At the boundary of the star $x=\mathcal{R}$, we match the solution \eqref{33} to the metric \eqref{2} to obtain
\begin{eqnarray}\label{41}
\begin{split}
\frac{A^{2}\left[(C\mathcal{R}^{2})^{3/2}+c_{1}C\mathcal{R}^{2}-c_{2}\sqrt{C^{2}\mathcal{R}^{4}(C\mathcal{R}^{2}+c_{1}\sqrt{C\mathcal{R}^{2}})}\right]^{2}}{C^{2}\mathcal{R}^{4}(C\mathcal{R}^{2}+c_{1}\sqrt{C\mathcal{R}^{2}})}=&\;1-\frac{2M}{\mathcal{R}}+\left(\frac{Q}{\mathcal{R}}\right)^{2},\\
\frac{1}{4C^{2}\mathcal{R}^{2}+4c_{1}C\sqrt{C\mathcal{R}^{2}}}=&\;\left[1-\frac{2M}{\mathcal{R}}+\left(\frac{Q}{\mathcal{R}}\right)^{2}\right]^{-1}.
\end{split}
\end{eqnarray}
Solving \eqref{41} for $c_{1}$ and $c_{2}$, we obtain
\begin{eqnarray}
\begin{split}
c_{1}=&\;\frac{Q^{2}+\mathcal{R}^{2}-4C^{2}\mathcal{R}^{4}-2M\mathcal{R}}{4C\mathcal{R}^{2}\sqrt{C\mathcal{R}^{2}}},\\
c_{2}=&\;\frac{A\sqrt{C\mathcal{R}^{2}(Q^{2}+\mathcal{R}^{2}-2M\mathcal{R})}\pm2\sqrt{C^{3}Q^{2}\mathcal{R}^{4}+C^{3}\mathcal{R}^{6}-2C^{3}M\mathcal{R}^{5}}}{CA\mathcal{R}^{2}\sqrt{C\mathcal{R}^{2}}}.
\end{split}
\end{eqnarray}
It is simply a matter of choosing the correct values of the constants to ensure that all physical conditions of positivity are met.

\subsection{Solution II}
For $\Delta=\frac{2Bx-C}{x}$, it is easy to verify that the potentials
\begin{eqnarray}\label{32}
\begin{split}
y(x)=&\;\frac{c_{2}x^{1/4}-4\sqrt{c_{1}}}{x^{1/4}},\\
z(x)=&\;\frac{c_{1}}{\sqrt{x}}
\end{split}
\end{eqnarray}
satisfy equation Theorem 1 above. Using the forms of the potentials $y$ and $z$ in \eqref{32}, we generate the exact solution to the field equations \eqref{10}
\begin{eqnarray}
\begin{split}
\text{e}^{2\nu}=&\;A^{2}\left(\frac{c_{2}x^{1/4}-4\sqrt{c_{1}}}{x^{1/4}}\right),\\
\text{e}^{2\lambda}=&\;\frac{\sqrt{x}}{c_{1}},\\
\rho=&\;\frac{3Cc_{1}^{3/2}}{(c_{2}x^{1/4}-4\sqrt{c_{1}})x^{3/2}}+\frac{4Bx^{3/2}+3Cc_{1}}{4x^{3/2}},\\
p_{\parallel}=&\;\frac{Cc_{1}^{3/2}}{(c_{2}x^{1/4}-4\sqrt{c_{1}})x^{3/2}}-\frac{4Bx^{3/2}-Cc_{1}}{4x^{3/2}},\\
p_{\perp}=&\;\frac{Cc_{1}^{3/2}}{(c_{2}x^{1/4}-4\sqrt{c_{1}})x^{3/2}}+\frac{4Bx^{3/2}-4C\sqrt{x}+Cc_{1}}{4x^{3/2}},\\
\Delta=&\;\frac{2Bx-C}{x},\\
E^{2}=&\;-\frac{6Cc_{1}^{3/2}}{(c_{2}x^{1/4}-4\sqrt{c_{1}})x^{3/2}}-\frac{4Bx^{3/2}-4C\sqrt{x}+3Cc_{1}}{2x^{3/2}},\\
m(x)=&\;\frac{6C\sqrt{x}-4Bx^{3/2}-3Cc_{1}}{3C^{3/2}}-\frac{4c_{1}^{3/2}}{\sqrt{C}(c_{2}x^{1/4}-4\sqrt{c_{1}})}.
\end{split}
\end{eqnarray}
Using \eqref{32} in \eqref{1}, the interior geometry is given by
\begin{eqnarray}\label{35}
\text{d}s^{2}_{-}=-\frac{A^{2}(c_{2}x^{1/4}-4\sqrt{c_{1}})^{2}}{\sqrt{x}}\;\text{d}t^{2}+\frac{\text{d}r^{2}}{4c_{1}C\sqrt{x}}+\frac{x}{C}(\text{d}\theta^{2}+\sin^{2}\theta\;\text{d}\phi^{2}).
\end{eqnarray}
At the boundary of the star $x=\mathcal{R}$, we match the solution \eqref{35} to the metric \eqref{2} to obtain
\begin{eqnarray}\label{38}
\begin{split}
\frac{A^{2}[c_{2}(C\mathcal{R}^{2})^{1/4}-4\sqrt{c_{1}}]^{2}}{\sqrt{C\mathcal{R}^{2}}}=&\;1-\frac{2M}{\mathcal{R}}+\left(\frac{Q}{\mathcal{R}}\right)^{2},\\
\frac{\mathcal{R}^{2}}{4c_{1}(C\mathcal{R}^{2})^{3/2}}=&\;\left[1-\frac{2M}{\mathcal{R}}+\left(\frac{Q}{\mathcal{R}}\right)^{2}\right]^{-1}.
\end{split}
\end{eqnarray}
Solving \eqref{38} for $c_{1}$ and $c_{2}$, we obtain
\begin{eqnarray}
\begin{split}
c_{1}=&\;\frac{Q^{2}+\mathcal{R}^{2}-2M\mathcal{R}}{4(C\mathcal{R}^{2})3/2},\\
c_{2}=&\;\frac{2(C\mathcal{R}^{2})^{7/4}}{C^{2}\mathcal{R}^{4}}\sqrt{\frac{\mathcal{R}^{2}+Q^{2}-2M\mathcal{R}}{(C\mathcal{R}^{2})^{3/2}}}\pm\frac{\sqrt{C^{4}\mathcal{R}^{8}+C^{4}Q^{2}\mathcal{R}^{6}-2C^{4}M\mathcal{R}^{7}}}{AC^{2}\mathcal{R}^{4}}.
\end{split}
\end{eqnarray}
It is simply a matter of choosing the correct values of the constants to ensure that all physical conditions of positivity are met.

%

\subsection{GENERALIZED REDUCTION}
We observe that for $\Delta=\frac{4Cx\dot{\mathcal{F}}\mathcal{F}+2Bx^{4}-5C\mathcal{F}^{2}-Cx^{3}}{x^{4}}$ the potentials
\begin{eqnarray}\label{45}
\begin{split}
y=&\;c_{1}\int\;\frac{\text{d}x}{\mid \mathcal{F}\mid}+c_{2},\\
z=&\;\frac{\mathcal{F}^{2}}{x^{3}},
\end{split}
\end{eqnarray}
where $\mathcal{F}=\mathcal{F}(x)\in \mathcal{C}^{\infty}$ is any arbitrary smooth and continuous function with the restrictions that its reciprocal is integrable and $\mathcal{F}\neq 0$. Using the forms of the potentials $y$ and $z$ in \eqref{45}, we generate the exact solution to the field equations \eqref{10}
\begin{eqnarray}\label{46}
\begin{split}
\text{e}^{2\nu}=&\;A^{2}\left[c_{1}\int\;\frac{\text{d}x}{\mid\mathcal{F}\mid}+c_{2}\right]^{2},\\
\text{e}^{2\lambda}=&\;\frac{x^{3}}{\mathcal{F}^{2}},\\
\rho=&\;\frac{3C\mathcal{F}^{2}}{2x^{4}}\left[\frac{2c_{1}x}{\mid\mathcal{F}\mid\left(c_{1}\int\frac{\text{d}x}{\mid\mathcal{F}\mid}+c_{2}\right)}+3\right]-\frac{3C\mathcal{F}\dot{\mathcal{F}}}{x^{3}}+B,\\
p_{\parallel}=&\;\frac{C\mathcal{F}^{2}}{2x^{4}}\left[\frac{2c_{1}x}{\mid\mathcal{F}\mid\left(c_{1}\int\frac{\text{d}x}{\mid\mathcal{F}\mid}+c_{2}\right)}+3\right]-\frac{C\mathcal{F}\dot{\mathcal{F}}}{x^{3}}-B,\\
p_{\perp}=&\frac{C\mathcal{F}^{2}}{2x^{4}}\left[\frac{2c_{1}x}{\mid\mathcal{F}\mid\left(c_{1}\int\frac{\text{d}x}{\mid\mathcal{F}\mid}+c_{2}\right)}-7\right]+\frac{3C\mathcal{F}\dot{\mathcal{F}}}{x^{3}}-\frac{C}{x}+B,\\
\Delta=&\;\frac{4Cx\dot{\mathcal{F}}\mathcal{F}+2Bx^{4}-5C\mathcal{F}^{2}-Cx^{3}}{x^{4}},\\
E^{2}=&\;\frac{C\mathcal{F}^{2}}{x^{4}}\left[1-\frac{6c_{1}x}{\mid\mathcal{F}\mid\left(c_{1}\int\frac{\text{d}x}{\mid\mathcal{F}\mid}+c_{2}\right)}\right]-\frac{2C\mathcal{F}\dot{\mathcal{F}}}{x^{3}}+\frac{2C}{x}-2B,\\
m(x)=&\;-\frac{8c_{1}B\mathcal{F}\dot{\mathcal{F}}\int\frac{\text{d}x}{\mathcal{F}}}{\sqrt{C}x^{3/2}\Phi}+\frac{4c_{1}\mathcal{F}\dot{\mathcal{F}}\int\frac{\text{d}x}{\mathcal{F}}}{\sqrt{Cx}\Phi}-\frac{2c_{1}x^{3/2}\int\frac{\text{d}x}{\mathcal{F}}}{\sqrt{C}\Phi}+\frac{8c_{1}B\mathcal{F}^{2}\int\frac{\text{d}x}{\mathcal{F}}}{\sqrt{C}x^{5/2}\Phi}\\
&-\frac{4c_{1}\mathcal{F}^{2}\int\frac{\text{d}x}{\mathcal{F}}}{\sqrt{C}x^{3/2}\Phi}+\frac{4c_{2}\mathcal{F}\dot{\mathcal{F}}}{\sqrt{Cx}\Phi}-\frac{2c_{2}x^{3/2}}{\sqrt{C}\Phi}+\frac{8c_{2}B\mathcal{F}^{2}}{\sqrt{C}x^{5/2}\Phi}-\frac{8c_{1}B\mathcal{F}^{2}}{\sqrt{C}x^{3/2}\mid\mathcal{F}\mid\Phi}\\
&+\frac{4c_{1}\mathcal{F}^{2}}{\sqrt{Cx}\mid\mathcal{F}\mid\Phi}-\frac{4c_{2}\mathcal{F}^{2}}{\sqrt{C}x^{3/2}\Phi}-\frac{8c_{2}B\mathcal{F}\dot{\mathcal{F}}}{\sqrt{C}x^{3/2}\Phi},
\end{split}
\end{eqnarray}
where 
\begin{eqnarray}
\Phi=\Phi(x)=2c_{1}B\int\frac{\text{d}x}{\mathcal{F}}-c_{1}x\int\frac{\text{d}x}{\mathcal{F}}-c_{2}x+2c_{2}B
\end{eqnarray}
We illustrate some of the solutions that are attainable from this generalized reduction in TABLE II below.
\begin{table}[h]
\centering	
\centering{\caption{Some models generated from the generalized reduction}}	
\begin{tabular}{|p{3cm}|p{7.5cm}|p{7.3cm}|}
\hline
\hline	
\centering{\textbf{Gravitational Potentials}} & \centering{\textbf{Exact Solution}} &$\;$\textbf{Matching at Stellar Boundary $x=\mathcal{R}$} \\
\hline
$y=\frac{c_{1}}{\sqrt{a}}x+c_{2}$&$\text{e}^{2\nu}=\frac{A^{2}(c_{1}x+c_{2}\sqrt{a})^{2}}{a}$ &$c_{1}=\frac{\pm\sqrt{a\mathcal{R}^{2}+aQ^{2}-2aM\mathcal{R}}-b\sqrt{a}A\mathcal{R}}{CA\mathcal{R}}$ \\
$z=\frac{a}{x^{3}}$ &$\text{e}^{2\lambda}=\frac{x^{3}}{a}$ &$c_{2}=b$ \\
&$\rho=\frac{3ac_{1}C}{x^{3}(c_{1}x+c_{2}\sqrt{a})}+\frac{2Bx^{4}+9aC}{2x^{4}}$ & \\
&$p_{\parallel}=\frac{ac_{1}C}{x^{3}(c_{1}x+c_{2}\sqrt{a})}-\frac{2Bx^{4}-3aC}{2x^{4}}$ & \\
&$p_{\perp}=\frac{ac_{1}C}{x^{3}(c_{1}x+c_{2}\sqrt{a})}-\frac{C(2x^{3}+7a)}{2x^{4}}+B$ & \\
&$\Delta=\frac{2Bx^{4}-Cx^{3}-5aC}{x^{4}}$ & \\
&$E^{2}=\frac{C(2x^{3}+a)}{x^{4}}-\frac{6ac_{1}C}{x^{3}(c_{1}x+c_{2}\sqrt{a})}-2B$ & \\
&$m(x)=\frac{2(6aC+3Cx^{3}-2Bx^{4})}{3C^{3/2}x^{5/2}}-\frac{4ac_{1}}{\sqrt{C}x^{3/2}(c_{1}x+c_{2}\sqrt{a})}$ & \\
\hline 
$y=\frac{2c_{1}\sqrt{\alpha x+\beta}}{\alpha}+c_{2}$ &$\text{e}^{2\nu}=\frac{A^{2}(2c_{1}\sqrt{\alpha x+\beta}+\alpha c_{2})^{2}}{\alpha^{2}}$ &$c_{1}=\frac{\pm\alpha\sqrt{\alpha CQ^{2}\mathcal{R}^{2}+\alpha C\mathcal{R}^{4}+\beta Q^{2}-2\alpha CM\mathcal{R}^{3}-2\beta M\mathcal{R}}}{A\mathcal{R}(2\alpha C\mathcal{R}^{2}+\beta)}$ \\
$z=\frac{\alpha x+\beta}{x^{3}}$ &$\text{e}^{2\lambda}=\frac{x^{3}}{\alpha x+\beta}$ &$\hspace{0.9cm}-\frac{b\alpha}{\sqrt{\alpha C\mathcal{R}^{2}+\beta}}$ \\
&$\rho=\frac{3\alpha c_{1}C\sqrt{\alpha x+\beta}}{x^{3}(2c_{1}\sqrt{\alpha x+\beta}+\alpha c_{2})}+\frac{3C(2\alpha x+3\beta)}{2x^{4}}+B$ & \\
&$p_{\parallel}=\frac{\alpha c_{1}C\sqrt{\alpha x+\beta}}{x^{3}(2c_{1}\sqrt{\alpha x+\beta}+\alpha c_{2})}+\frac{C(2\alpha x+3\beta)}{2x^{4}}-B$ &$c_{2}=b$ \\
&$p_{\perp}=\frac{\alpha c_{1}C\sqrt{\alpha x+\beta}}{x^{3}(2c_{1}\sqrt{\alpha x+\beta}+\alpha c_{2})}-\frac{C(2x^{3}+4\alpha x+7\beta)}{2x^{4}}+B$ & \\
&$\Delta=\frac{2Bx^{4}-Cx^{3}-3\alpha Cx-5\beta C}{x^{4}}$ & \\
&$E^{2}=\frac{\beta C}{x^{4}}+\frac{2C}{x}-\frac{6\alpha c_{1}C\sqrt{\alpha x+\beta}}{x^{3}(2c_{1}\sqrt{\alpha x+\beta}+\alpha c_{2})}-2B$ & \\
&$m(x)=\frac{2\alpha}{\sqrt{C}x^{3/2}}+\frac{4\beta}{\sqrt{C}x^{5/2}}+2\sqrt{\frac{x}{C}}-\frac{4B}{3}\left(\frac{x}{C}\right)^{3/2}$ & \\
&$\hspace{1.4cm}-\frac{4\alpha c_{1}\sqrt{\alpha x+\beta}}{\sqrt{C}x^{3/2}(2c_{1}\sqrt{\alpha x+\beta}+\alpha c_{2})}$ & \\
\hline
\end{tabular}
\end{table}


\newpage
\section{Discussion}\label{discussion}
In this paper, we have systematically investigated the Einstein-Maxwell equations for quark stars that have a linear equation of state with the MIT Bag Model constants \cite{Chodus} and \cite{DeGrand}. Cognate studies of this nature have been conducted in great detail and are therefore non-contemporary; Sunzu \textit{et. al.} \cite{SunzuMaharajRay2} and \cite{SunzuMaharajRay1} and Maharaj \textit{et. al.} \cite{SunzuMaharajRay3}, for example,  have conducted latter-day studies in this area in great detail and produced new solutions to the field equations. However, the application of Lie symmety analysis to these equations for the particular situation of the MIT Bag Model constants is new. Lie symmetry analysis to the master gravitational equation generates a new class of solutions that are interesting. We have identified four particular cases for the choice of the measure of the anisotropy that produced a relationship between the two gravitational potentials. For the most general case of the measure of the anisotropy, where it is allowed to vary arbitrarily with respect to the radial coordinate, we have postulated a general correspondence between the two gravitational potentials. Using the result, we generated two exact solutions for specific choices of the measure of the anisotropy and demonstrated the physical viability of these solutions by matching the solutions to the exterior Reissner-Nordstr$\ddot{\text{o}}$m metric at the stellar boundary. For the variegated solution sets we are able to generate closed form results for gravitational potentials and were therefore able to express all physical parameters energy density, radial pressure, tangential pressure, measure of anisotropy, electric field intensity and the mass of the gravitating hypersphere. Wherefore, a physical analysis can be performed to obtain feasible and practicable solutions to the Einstein-Maxwell equations.  We take note that the models generated in \S VI are for core envelope stars; that is, stellar hyperspherical structures with singularities at their centres. We believe that it is possible to attain solutions that are singularity-free for a suitable choice of the measure of the anisotropy $\Delta$ as per the investigations carried out by previous researchers; a definitive example being the results obtained by Sunzu \textit{et. al.} \cite{SunzuMaharajRay2} and \cite{SunzuMaharajRay1} and Maharaj \textit{et. al.} \cite{SunzuMaharajRay3} wherefore they found quark star models with $\{\rho, p_{\parallel}, p_{\perp}, \Delta, E^{2}\}$ obeying positivity relations and corresponding stellar candidates from observational cosmology.

A useful consideration would be the more general linear equation of state $p_{\parallel}=\alpha\rho-\beta$ with the underlying gravitational equation 
\begin{eqnarray} \label{48}
\begin{split}
&4(\alpha+1)Cx^{2}z\ddot{y}+(2\alpha Cx^{2}\dot{z}+2Cx^{2}\dot{z}+8Cxz)\dot{y}\\
&+[5\alpha Cx\dot{z}+Cx\dot{z}+\alpha Cz+Cz+2\beta x-\alpha x\Delta(x)-x\Delta(x)-\alpha C-C]y=0.
\end{split}
\end{eqnarray}
Obtaining a correspondence between the gravitational potentials in \eqref{48} would prove a more utilitarian appraisal and would be able to generalize the results obtained in this paper.


\begin{acknowledgements}
MAZK and RBN thank the  University of KwaZulu-Natal of the Republic of South Africa for making this research possible.
\end{acknowledgements}


\end{document}